\documentclass[3p,times]{elsarticle}

\usepackage{ecrc}

\usepackage{epstopdf}

\usepackage{natbib}

\volume{00}

\firstpage{1}

\journalname{Nuclear Physics A}

\runauth{Roy A. Lacey}

\jid{nupha}

\jnltitlelogo{Nuclear Physics A}

\usepackage{graphicx}
\usepackage{amsmath,amssymb,amsfonts}
 \usepackage{lineno}
\newcommand{\be}{\begin{eqnarray}}
\newcommand{\ee}{\end{eqnarray}}

\newcommand{\sqsn}{\mbox{$\sqrt{s_{_{NN}}}$}}

\begin{document}

\begin{frontmatter}



\title{Beam energy dependence of the expansion dynamics in relativistic heavy ion collisions: Indications for 
the critical end point?}

%% Single author (and collaboration) - please insert
\author{Roy A. Lacey}
\fntext[col1] {Roy.Lacey@Stonybrook.edu}
\address{Depts. of Chemistry \& Physics, Stony Brook University, NY 11974}

%% For multiple authors, replace the above by:

%\author[label1]{Author1}
%\author[label2]{Author2}

%\address[label1]{Address 1}
%\address[label2]{Address 2}

\begin{abstract}

The flow harmonic $v_{n}$ and the emission source radii $R_{\text{out}}$, 
$R_{\text{side}}$ and $R_{\text{long}}$ are studied for a broad range of 
centrality selections and beam collision energies in Au+Au ($\sqrt{s_{NN}}= 7.7 - 200$ GeV) 
and Pb+Pb ($\sqrt{s_{NN}}= 2.76$ TeV) collisions at RHIC and the LHC respectively. 
They validate the acoustic scaling patterns expected for hydrodynamic-like expansion over the entire range 
of beam energies studied. The combined data sets allow estimates for the \sqsn\ dependence of the mean expansion 
speed $\left<c_s\right>$, emission duration $\left<\Delta\tau\right>$ and the viscous coefficients 
$\left<\beta''\right>$ that encode the magnitude of  the specific shear viscosity $\left<\eta/s\right>$.
The estimates indicate initial-state model independent values of $\left<\eta/s\right>$ which are larger 
for the plasma produced at 2.76 TeV (LHC) compared to that produced at 200 GeV (RHIC) 
($\left<4\pi\eta/s\right>_{\text{LHC}}=2.2\pm 0.2$ and $\left<4\pi\eta/s\right>_{\text{RHIC}}=1.3\pm 0.2$).
They also show a non-monotonic \sqsn\  dependence for $\left<\beta''\right>$, $\left<c_s\right>$ and $\left<\Delta\tau\right>$,
with minima for $\left<\beta''\right>$ and $\left<c_s\right>$, and a complimentary maximum for $\left<\Delta\tau\right>$. 
These dependencies signal a significant change in reaction dynamics in a narrow span of  $\sqsn$, which may be 
linked to reaction trajectories close to the critical end point (CEP) in the phase diagram for nuclear matter. 
%
%A template for preparing contributions to the proceedings of Quark Matter 2014. The file should be compiled with {\em pdflatex}. 
%Figures can be pdf or eps, as illustrated in Fig.~\ref{fig:generic}. If the conversion eps $\to$ pdf does not work automatically on 
%your system, you can convert eps files to pdf using a tool like {\em epstopdf}. 
%For special options see\\ \verb!http://www.elsevier.com/author-schemas/preparing-crc-journal-articles-with-latex!.
%
\end{abstract}


\end{frontmatter}
% \linenumbers
%% \begin{linenumbers}, end it with \end{linenumbers}. Or switch it on
% \linenumbers


\section{Introduction}
\label{intro}

Heavy ion collisions provide an important avenue for studying the phase diagram for 
Quantum Chromodynamics (QCD) \cite{Itoh:1970,Shuryak:1983zb,Stephanov:1998dy}. 
The location of the phase boundaries and the critical end point (CEP), in the plane of temperature vs. baryon chemical 
potential [($T,\mu_B$)-plane], are fundamental ``landmarks'' of this phase diagram \cite{Asakawa:1989bq}. 
Lattice QCD calculations suggest that the quark-hadron transition is a 
crossover at high $T$ and small $\mu_{B}$ or high collision energies (\sqsn)~\cite{Aoki:2006we}. 
Experimental results from the Relativistic Heavy Ion Collider (RHIC) at \sqsn = 200~GeV and the Large Hadron Collider (LHC) 
at \sqsn = 2.76~TeV, indicate that this transition results in the production of a strongly coupled plasma of de-confined quarks and 
gluons (sQGP) with low specific shear viscosity ${\eta}/{s}$, {\em i.e.} the ratio of shear viscosity
$\eta$ to entropy density $s$ \cite{Lacey:2006bc}. The validation of this crossover transition, which  
is a necessary, albeit insufficient, requirement for the existence of the CEP, serves as an important 
impetus for the ongoing experimental searches.

A current strategy for (i) establishing the essential ``landmarks'' of the phase diagram and 
(ii) pinning down the thermodynamic and transport properties of each QCD phase, is 
centered on measurements in energy scans designed to access the broadest possible $(T,\mu_B)$-domain of  
the phase diagram. In this proceedings we follow this lead by leveraging the combined measurements for 
anisotropic flow and  HBT radii, recently obtained  by PHENIX and STAR in the first RHIC Beam Energy Scan (BES-I) 
and by ATLAS, ALICE and CMS at the LHC.

\section{Probes for transport properties and the Critical End Point (CEP)}

The expansion dynamics of  relativistic heavy ion collisions is strongly influenced 
by the transport properties of the created medium, as well as the path of the 
reaction trajectory in the  ($T, \mu_{B}$)-plane. Such an influence can manifest as quantifiable 
changes in the magnitude of the space-time extent of the emission source, characterized by the so-called  HBT
radii $R_{\text{out}}$, $R_{\text{side}}$ and $R_{\text{long}}$;  
the square of (i) the emission source lifetime $\tau^2 \propto R_{\text{long}}^2$, 
(ii) its geometric size $R_{\text{geo}}^2 \propto R_{\text{side}}^2$
and (iii) the emission duration $\Delta \tau^2 \propto (R_{\text{out}}^2 - R_{\text{side}}^2) $ \cite{Chapman:1994yv}.
It can also manifest as a significant modulation of  the anisotropic flow coefficients $v_n$, 
depending on the magnitude of $\eta/s$.
The LHC measurements at $\sqrt{s_{NN}}= 2.76$ TeV, allow investigations of the space-time extent 
and $\eta/s$ at high $T$ and small $\mu_{B}$; they compliment similar measurements from BES-I which 
allow a systematic study  for the $\mu_B$ and $T$ values spanned by the collision 
energy range $\sqrt{s_{NN}}= 7.7 - 200$ GeV. Here, it is noteworthy that currently,
there are only a few experimental constraints for the ($T,\mu_B$)-dependence of $\eta/s$, especially at the lower 
beam energies \cite{Lacey:2013qua}. 

At the CEP or close to it, anomalies in the dynamic properties of the medium can drive abrupt  
changes in transport coefficients and relaxation rates to give a non-montonic dependence of  
$\frac{\eta}{s}(T, \mu_B)$ \cite{Lacey:2006bc,Csernai:2006zz,Lacey:2007na}. 
An emitting system produced in the vicinity of the CEP is also expected to show a stalling of the 
expansion speed and a larger emission duration  manifested as a difference between $R_{\text{out}}$ and $R_{\text{side}}$
($\Delta \tau^2 \propto (R_{\text{out}}^2 - R_{\text{side}}^2) $) \cite{Rischke:1996em}.
Here, the rationale is that, in the vicinity of the CEP, the equation of state (EOS) ``softens" considerably
and this slows down the speed of expansion and prolongs the emission duration 
to give $R_{\text{out}}>R_{\text{side}}$. 

In prior work \cite{Lacey:2013is,Lacey:2011ug}, we have used the participant eccentricities ($\varepsilon_n$) 
and initial transverse size $\bar{R}$ (${1}/{\bar{R}}=\sqrt{\left({1}/{\sigma_x^2}+{1}/{\sigma_y^2}\right)}$, 
where $\sigma_x$ and $\sigma_y$ are the respective root-mean-square widths of the density distributions)
obtained with Monte Carlo Glauber (MC-Glauber) simulations, to validated the acoustic nature of 
of the expansion dynamics  \cite{Staig:2010pn} in RHIC and LHC collisions. This acoustic property 
predicts a linear relationship between the expansion time ($t$) and the initial traverse size ($t \propto \bar{R}$), 
as well as a characteristic linear dependence of $\ln(v_n/\varepsilon_n)$ on 
both $n^2$ and $1/\bar{R}$, with slopes $\beta'\propto (\eta/s)$ and $\beta''\propto (\eta/s)$.
We use the latter scaling patterns in conjunction with viscous hydrodynamical 
calculations \cite{Song:2011hk,cms_ulc_note}, to calibrate $\beta'\text{ and }\beta''$
and make estimates of $\left<\eta/s\right>$ for the plasma produced in Au+Au and Pb+Pb 
collisions at 200 GeV and 2.76 TeV respectively. A further study of the \sqsn\ dependence of 
$\beta''$, $c_s$ and $\Delta\tau$ is then used to search for non-monotonic patterns 
which could signal the presence of the CEP.

\section{Results}
%
%%%%%%%%%%%%%%%%%%%%%%%%%%%%%%%%%  Figure 1  %%%%%%%%%%%%%%%%%%%%%%%%%%%%%%%%
%
\begin{figure}[t]
\begin{center}
\includegraphics[width=0.8\textwidth]{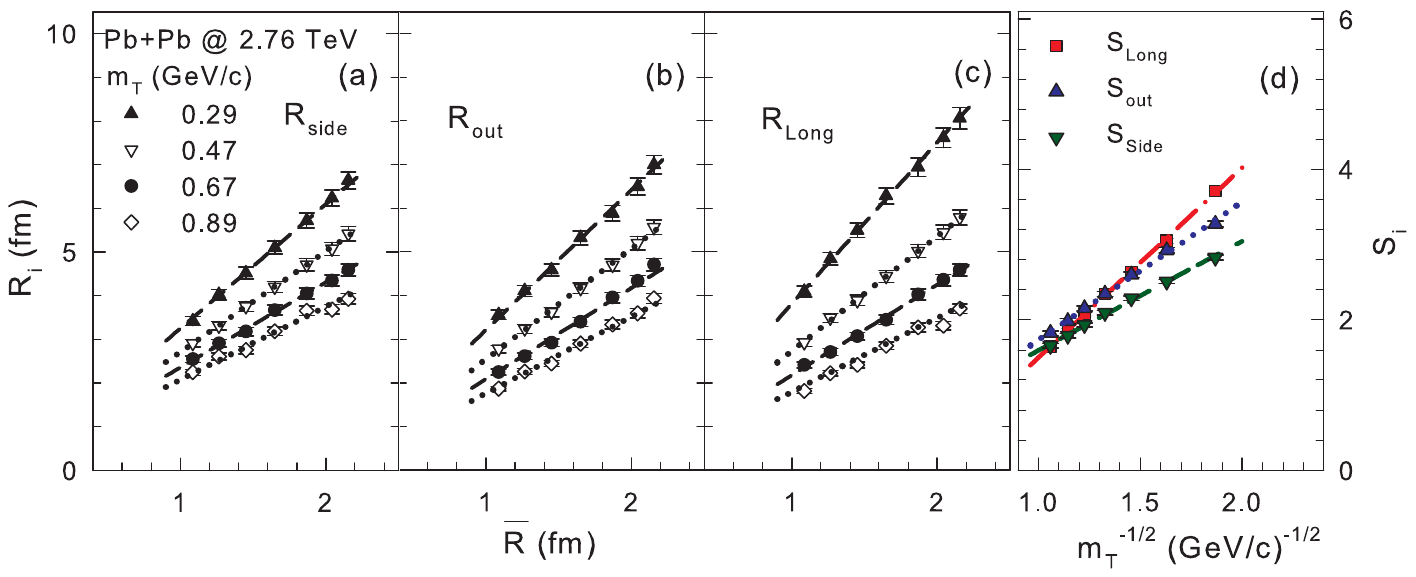}
\end{center}
\vspace{-0.75cm}
\caption{(Color online) HBT radii vs. $\bar{R}$  for several $m_T$ cuts (as indicated) for (a) $R_{\text{side}}$, 
(b) $R_{\text{out}}$ and (c) $R_{\text{long}}$ for Pb+Pb collisions at \sqsn =2.76 TeV; the data are 
taken from Ref.~\cite{Kisiel:2011jt}. The dashed and dotted curves indicate linear fits to the data (see text). 
(d) $S_i$ vs. $1/\sqrt{m_T}$;  $S_i$ are slopes obtained from the respective linear fits  to 
the scaled values of  $R_{\text{side}}$, $R_{\text{out}}$ and $R_{\text{long}}$, shown in (a), (b) and (c).
The dashed, dashed-dot and dotted curves in this panel, represent linear fits.
}
\label{Fig1}
\vspace{-4pt}
\end{figure}
%
%
%%%%%%%%%%%%%%%%%%%%%%%%%%%%%%%%  Figure 2  %%%%%%%%%%%%%%%%%%%%%%%%%%%%%%%%
\begin{figure}[t]
\includegraphics[width=1.0\textwidth]{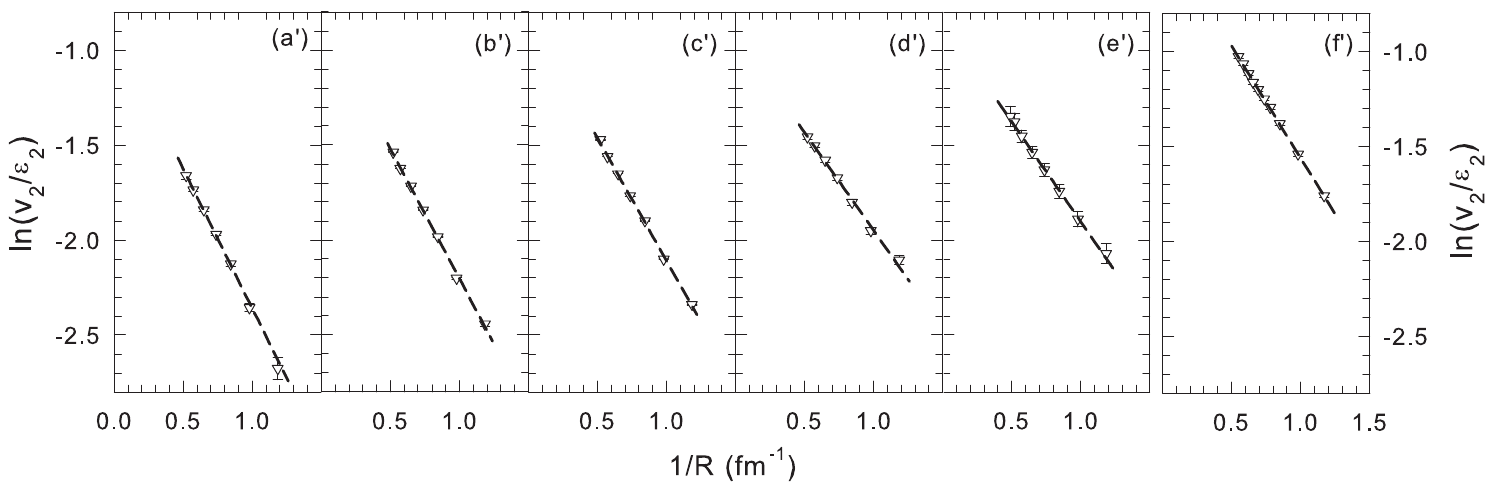}  
\caption{(($\text{a}')-(\text{e}'$)) $\ln(v_2/\varepsilon_2)$ vs. $1/\bar{R}$ for $p_T$-integrated $v_{2}$ ($p_T \gtrsim 0.2$ GeV/c)
for Au+Au collisions. ($\text{f}'$) $\ln(v_2/\varepsilon_2)$ vs. $1/\bar{R}$ for $p_T$-integrated $v_{2}$ ($p_T= 0.3-3$ GeV/c)
for Pb+Pb collisions.
The data for Au+Au and Pb+Pb collisions are taken from Refs. \cite{Adamczyk:2012ku,Agakishiev:2011eq}
and Ref. \cite{Chatrchyan:2012ta} respectively. The dashed curves represent linear fits to the data; error bars are statistical only.
}
\label{Fig2}
\end{figure}

Representative summaries of the the scaling properties of the HBT radii and $v_n$ are shown 
in Figs.~\ref{Fig1} and  \ref{Fig2} respectively. 
Figs.~\ref{Fig1}(a), (b) and (c)  validate the expected linear dependence of  $R_{\text{side}}$, $R_{\text{out}}$ 
and $R_{\text{long}}$ on $\bar{R}$. They also show the expected  decrease in the slope of the
respective scaling curves  (for $R_{\text{side}}$, $R_{\text{out}}$ and $R_{\text{long}}$) with increasing 
transverse mass $m_T$.
The latter confirms the important influence of  the space-momentum correlations which result from 
collective expansion.  Fig. \ref{Fig1}(d) shows that the slopes $S_i$, obtained from linear fits 
(dashed and dotted curves) to the data in panels (a)-(c), scale as $1/\sqrt{m_T}$ and the space-momentum 
correlations are largest (smallest) in the long (side) direction. They also indicate that, for a given \sqsn, the full 
set of differential measurements (as a function of centrality and $m_T$) for each radius, can be made to 
scale to a single curve. 
Similar scaling patterns were observed for the full range of \sqsn\ values spanned by the 
PHENIX and STAR data sets.

Figures~\ref{Fig2}($\text{a}')-(\text{f}'$) show the expected linear dependence 
of $\ln(v_n/\varepsilon_n)$ vs. $1/\bar{R}$ for the full range of available 
beam energies. This pervasive pattern of scaling provides the basis for a consistent method of 
extraction of the viscous coefficient $\beta^{''} \propto \eta/s$, via linear fits to the scaled data 
for each beam energy. The characteristic linear dependence of $\ln(v_n/\varepsilon_n)$ on 
both $n^2$ and $1/\bar{R}$, with slopes $\beta'\propto (\eta/s)$ and $\beta''\propto (\eta/s)$
was also observed for viscous hydrodynamical calculations \cite{Lacey:2013eia}. 
Consequently, such calculations were used to calibrate $\beta'$ and $\beta''$ and 
extract $\eta/s$ for the plasma produced in Au+Au and Pb+Pb collisions at 
200 GeV (RHIC) and 2.76 TeV (LHC) respectively. This procedure give the 
values $\left<4\pi\eta/s\right>_{\text{RHIC}}=1.3\pm 0.2$) and 
$\left<4\pi\eta/s\right>_{\text{LHC}}=2.2\pm 0.2$ which are insensitive to 
the initial-state geometry model employed \cite{Lacey:2013eia}.

\begin{figure}
\begin{minipage}{.44\textwidth}
  \includegraphics[width=1.0\textwidth]{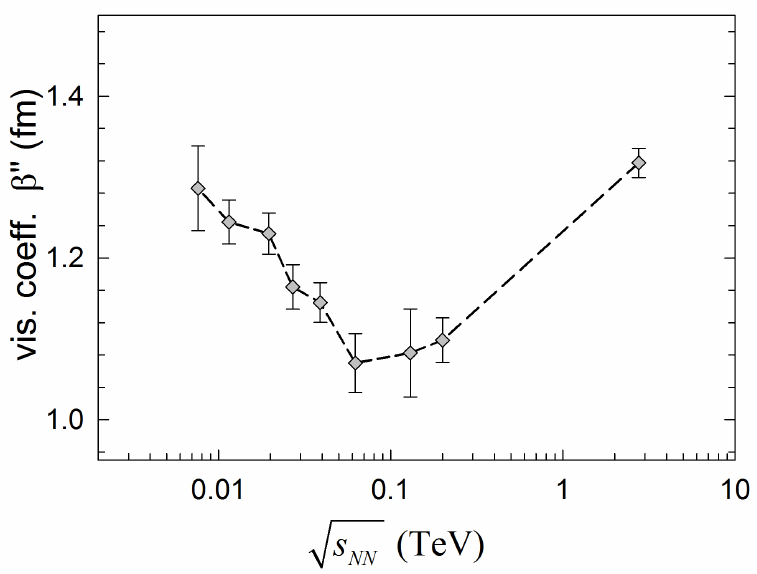}
  \caption{Viscous coefficient $\beta''$ vs. $\sqrt{s_{NN}}$, extracted from linear 
fits to $\ln(v_2/\varepsilon_2)$ vs. $1/\bar{R}$; error bars are statistical only. 
The dashed curve is drawn to guide the eye.
}
  \label{Fig5}
\end{minipage}
\begin{minipage}{.52\textwidth}
  \includegraphics[width=1.0\textwidth]{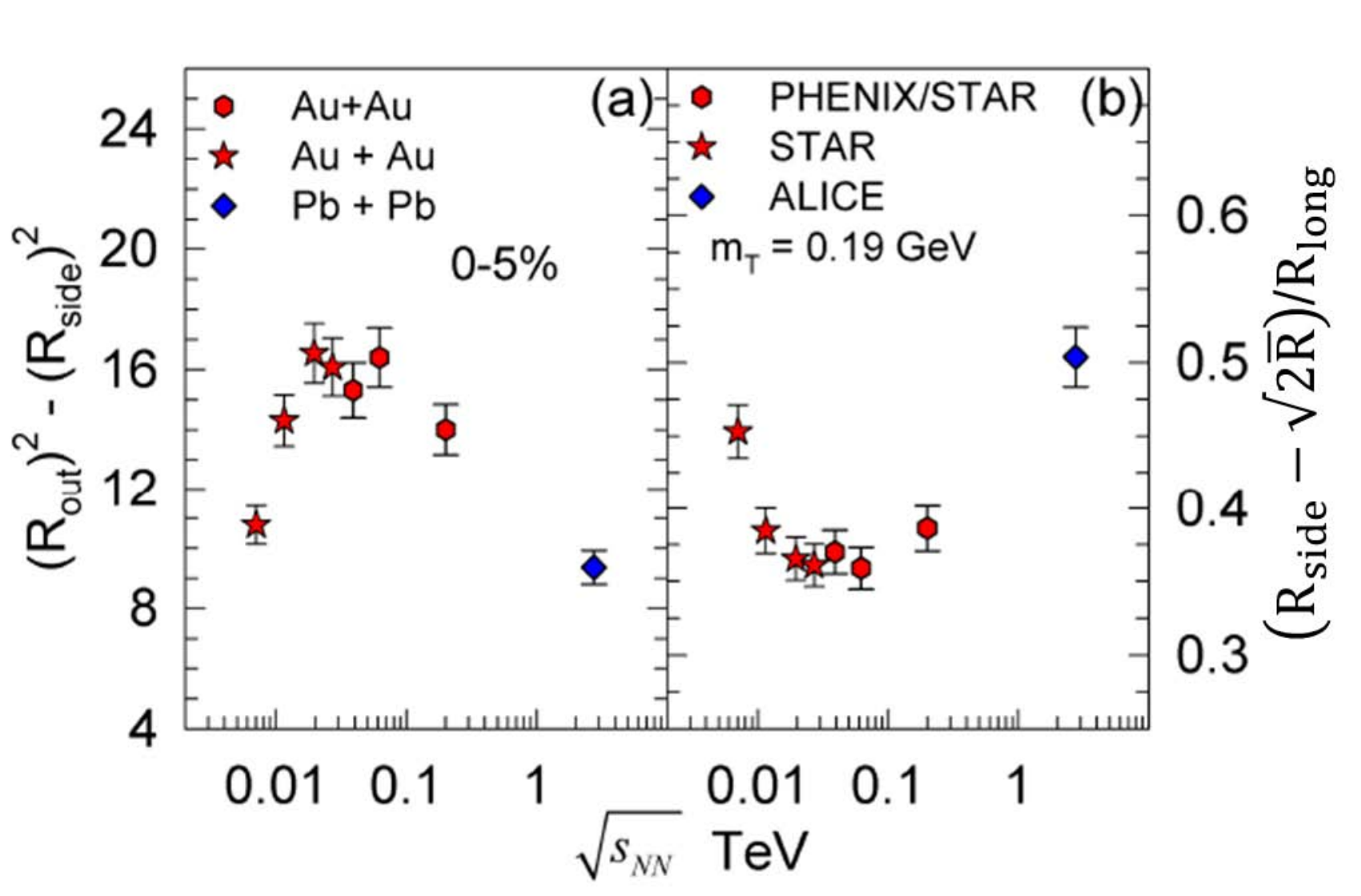}
  \caption{(Color online)  \sqsn\ dependence of  (a) $ (R_{\text{out}}^2 - R_{\text{side}}^2) \propto \Delta \tau^2 $, 
(b) [($R_{\text{side}}- \sqrt{2}\bar{R}$)/$R_{\text{long}}$] $\propto \left<c_s\right>$. 
The HBT radii used for these extractions are taken from preliminary PHENIX data 
and  Refs. \cite{Kisiel:2011jt,Adamczyk:2014mxp}.
}
  \label{Fig6}
\end{minipage}
\end{figure}

 The results from a search for possible non-monotonic patterns linked to the CEP, are 
summarized in Figs. \ref{Fig5} and  \ref{Fig6}. The \sqsn\  dependence for $\beta''$ (Fig. \ref{Fig5})
shows a decreasing trend from $7.7$~GeV to approximately 62.4~GeV, followed by a relatively 
slow increase from $\sqrt{s_{NN}}= 62.4$~GeV - 2.76~TeV. 
Here, it should be emphasized that the error bars for the extractions made at 62.4, 130 and 200 GeV, as well as a  
lack of measurements between 39 and 62.4 GeV, do not allow a definitive estimate of the 
precise location of this apparent minimum. Nonetheless, we associate this characteristic \sqsn\ dependence of $\beta^{''}$  
with the expected trend of $\frac{\eta}{s}(T,\mu_B)$  for reaction 
trajectories in the vicinity of the CEP  \cite{Csernai:2006zz,Lacey:2007na}.  If this is so, 
such trajectories should  also lead to signatures indicative of  a softening of the EOS and  
a prolonged emission duration. The results from the tests for such signatures are shown in 
Fig.  \ref{Fig6}.

The \sqsn\ dependence of  $ (R_{\text{out}}^2 - R_{\text{side}}^2) \propto \Delta \tau^2 $ and 
($R_{\text{side}}- \sqrt{2}\bar{R}$)/$R_{\text{long}}$ are shown in 
Figs.  \ref{Fig6} (a) and (b) respectively. ($R_{\text{side}}- \sqrt{2}\bar{R}$)/$R_{\text{long}}$ is 
used as a proxy for the expansion speed $c_s$ since ($R_{\text{side}}- \sqrt{2}\bar{R}$) represent the 
expansion radius and  $R_{\text{long}} \propto \tau$. Fig.~\ref{Fig6} shows the anticipated 
non-monotonic trends; they indicate a maximum for $\left<\Delta\tau\right>$ and a complimentary minimum 
for $\left<c_s\right>$ in the same narrow span of $\sqsn$. These dependencies signal a significant change in reaction 
dynamics which could also be linked to reaction trajectories close to the CEP.

%
%%%%%%%%%%%%%%%%%%%%%%%%%%%%%%%% N Figure 3  %%%%%%%%%%%%%%%%%%%%%%%%%%%%%%%%

%
%%%%%%%%%%%%%%%%%%%%%%%%%%%%%%%%%%  Figure 1  %%%%%%%%%%%%%%%%%%%%%%%%%%%%%%%%
%%
%\begin{figure}[t]
%%\begin{center}
%\includegraphics[width=1.0\linewidth]{Figs/Rbar_scaling_alice-b.pdf}
%%\end{center}
%\vspace{-0.75cm}
%\caption{(Color online) HBT radii vs. $\bar{R}$  for several $m_T$ cuts (as indicated) for (a) $R_{\text{side}}$, 
%(b) $R_{\text{out}}$ and (c) $R_{\text{long}}$ for Pb+Pb collisions at \sqsn =2.76 TeV; the data are 
%taken from Ref.~\cite{Kisiel:2011jt}. The dashed and dotted curves indicate linear fits to the data (see text). 
%(d) $S_i$ vs. $1/\sqrt{m_T}$;  $S_i$ are slopes obtained from the respective linear fits  to 
%the scaled values of  $R_{\text{side}}$, $R_{\text{out}}$ and $R_{\text{long}}$, shown in (a), (b) and (c).
%The dashed, dashed-dot and dotted curves in this panel, represent linear fits.
%}
%\label{fig3}
%\vspace{-4pt}
%\end{figure}
%%

%%%%%%%%%%%%%%%%%%%%%%%%%%%%%  Summary  %%%%%%%%%%%%%%%%%%%%%%%%%%%%%%

%
\section{Conclusions}

In summary, we have presented a detailed study of  the expansion dynamics in relativistic heavy ion 
collisions, using the combined data sets for flow and HBT measurements in Pb+Pb collisions at $\sqrt{s_{NN}}= 2.76$ TeV 
and  Au+Au collisions spanning $\sqrt{s_{NN}}= 7.7 - 200$~GeV. 
Our study shows that the scaling properties of these measurements 
validate the characteristic signatures expected for sound propagation in the 
matter produced in these collisions.  They also allow estimates for the \sqsn\ dependence of the mean expansion 
speed $\left<c_s\right>$, emission duration $\left<\Delta\tau\right>$ and the viscous coefficients 
$\left<\beta''\right>$ that encode the magnitude of  the specific shear viscosity $\left<\eta/s\right>$.
The estimates indicate a larger value of $\left<\eta/s\right>$ for the plasma produced at 2.76 TeV (LHC) compared 
to that produced at 200 GeV (RHIC) 
($\left<4\pi\eta/s\right>_{\text{LHC}}=2.2\pm 0.2$ and $\left<4\pi\eta/s\right>_{\text{RHIC}}=1.3\pm 0.2$);
these values are insensitive to the choice of the initial-state model employed for the extractions.
They also show a non-monotonic \sqsn\  dependence for $\left<\beta''\right>$, $\left<c_s\right>$ and $\left<\Delta\tau\right>$,
with minima for $\left<\beta''\right>$ and $\left<c_s\right>$, and a complimentary maximum for $\left<\Delta\tau\right>$. 
These dependencies signal an important change in the reaction dynamics for a narrow range of  $\sqsn$, which may be 
linked to reaction trajectories close to the critical end point. Further detailed studies, with improved errors and modeling, 
are required to make a more precise mapping, as well as to confirm if the observed patterns 
for $\left<\beta''\right>$, $\left<c_s\right>$ and $\left<\Delta\tau\right>$, are definitively linked to
decay trajectories close to the critical end point in the phase diagram for nuclear matter.

%%%%%%%%%%%%%%%%%%%%%%%%%%  Acknowledgements  %%%%%%%%%%%%%%%%%%%%%%%%%%
\section*{Acknowledgments}
%{\bf Acknowledgments:}
%%%%%%%%%%%%%%%%%%%%%%%%%%%%%%%%%%%%%%%%%%%%%%%%%%%%%%%%%%%%%%%%%%%%%%%%%%%%%%%%%%%%%%%%%%%%%%%%%%%%%%%%%%%%%%%%%%%%%%%%
This research is supported by the US DOE under contract DE-FG02-87ER40331.A008.
%and by the NSF under award number PHY-1019387.
 
%%%%%%%%%%%%%%%%%%%%%%%%%%%%%%%%%%%%%%%%%%%%%%%%%%%%%%%%%%%%%%%%%%%%%%%%%%%%%%%%%%%%%%%%%%%%%%%%%%%%%%%%%%%%%%%%%%%%%%%%%

%%%%%%%%%%%%%%%%%%%%%%%%%%%%%  References  %%%%%%%%%%%%%%%%%%%%%%%%%%%%%%

%
\bibliography{viscous_coefficient}

\end{document}